\documentclass[epj,twocolumn]{webofc}
\usepackage[varg]{txfonts}   
\usepackage{aas_macros}
\woctitle{The Innermost Regions of Relativistic Jets and Their Magnetic Fields}
\begin{document}
\title{Intrinsic brightness temperatures of blazar jets at 15 GHz}
%
%

\author{Talvikki Hovatta\inst{1}\fnsep\thanks{\email{thovatta@caltech.edu}} \and
        Erik M. Leitch\inst{1,2} \and
        Daniel C. Homan\inst{3} \and
      Kaj Wiik\inst{4} \and
    Matthew L. Lister\inst{5} \and
  Walter Max-Moerbeck\inst{6} \and
 Joseph, L. Richards\inst{5} \and
 Anthony C. S. Readhead\inst{1}}

\institute{California Institute of Technology, Pasadena, CA, USA
\and
           University of Chicago, Chicago, IL, USA
\and
           Denison University, Granville, OH, USA
\and
University of Turku, Piikki\"o, Finland
\and 
Purdue University, West Lafayette, IN, USA
\and
National Radio Astronomy Observatory, Socorro, NM, USA.
          }

\abstract{
We have developed a new Bayesian Markov Chain Monte Carlo method to deconvolve light curves of blazars into individual flares, including proper estimation of the fit errors. We use the method to fit 15\,GHz light curves obtained within the OVRO 40-m blazar monitoring program where a large number of AGN have been monitored since 2008 in 
 support of the Fermi Gamma-Ray Space Telescope mission. The time scales obtained from the fitted models are used to calculate the variability brightness temperature of the sources. Additionally, we have calculated brightness temperatures of a sample of these objects using Very Long Baseline Array data from the MOJAVE survey. Combining these two data sets enables us to study the intrinsic brightness temperature distribution in these blazars at 15 GHz. Our preliminary results indicate that the mean intrinsic brightness temperature in a sample of 14 sources is near the equipartition brightness temperature of $\sim 10^{11}$K.
}
\maketitle
\section{Introduction}\label{intro}
Blazars are active galactic nuclei (AGN) whose jets are pointing close to the line of sight. The emission of blazars is highly beamed by relativistic effects that can be quantified by the Doppler boosting factor. The amount of Doppler boosting determines both the observed luminosity of the object and the time scale of variations, with more boosted objects appearing brighter and varying on shorter time scales. In order to compare the intrinsic properties of a blazar sample, the amount of Doppler boosting needs to be accounted for. One way to estimate the amount of Doppler boosting is to study the observed brightness temperatures of the jets and compare these to the intrinsic brightness temperature that can be either assumed based on theoretical arguments, or be calculated if the brightness temperature of the object can be determined in two independent ways.

Already four decades ago it was shown that there is a limit to the maximum intrinsic brightness temperature in the jets of AGN that is dictated by the inverse Compton catastrophe limit, $T_{\rm b,int} = 10^{12}$\,K \cite{kellermann69}. Any observed brightness temperatures above that limit are due to Doppler boosting. Later, based on arguments that the magnetic and particle energy density in the jets are in equipartition, it was shown that a more likely limit for the intrinsic brightness temperature is the equipartition brightness temperature $T_{\rm b,eq} \sim 10^{11}$\,K \cite{readhead94}.

The brightness temperature is defined as
\begin{equation}
T_{\rm b} \propto \frac{S}{\theta^2\nu^2},
\end{equation}
where S is the flux density of the object, $\theta$ is its angular diameter and $\nu$ is the observing frequency.
Doppler boosting affects the observed brightness temperatures in different ways depending on the way it is determined. In the case of very long baseline interferometry (VLBI) observations, where the angular diameter can be directly measured, the observed brightness temperature is defined as $T_{\rm VLBI} = \delta T_{\rm b,int}$, where $\delta$ is the Doppler boosting factor. Another way to determine the brightness temperature is by assuming that the rise time of a flare corresponds to the light-travel time across the source, in which case one can estimate the angular size based on the observed variability time scale.  The variability brightness temperature is defined as $T_{\rm var} = \delta^3 T_{\rm b,int}$ \cite{lahteenmaki99}. Thus, by estimating the observed brightness temperature using both VLBI and variability information, one can solve for the intrinsic brightness temperature
\begin{equation}\label{eq:Tint}
T_{\rm b,int} = \sqrt{\frac{T_{\rm VLBI}^3}{T_{\rm var}}}.
\end{equation}
Alternatively, if the intrinsic brightness temperature is known, one can use either observed brightness temperature to estimate the amount of Doppler boosting.
\begin{figure*}[!ht]
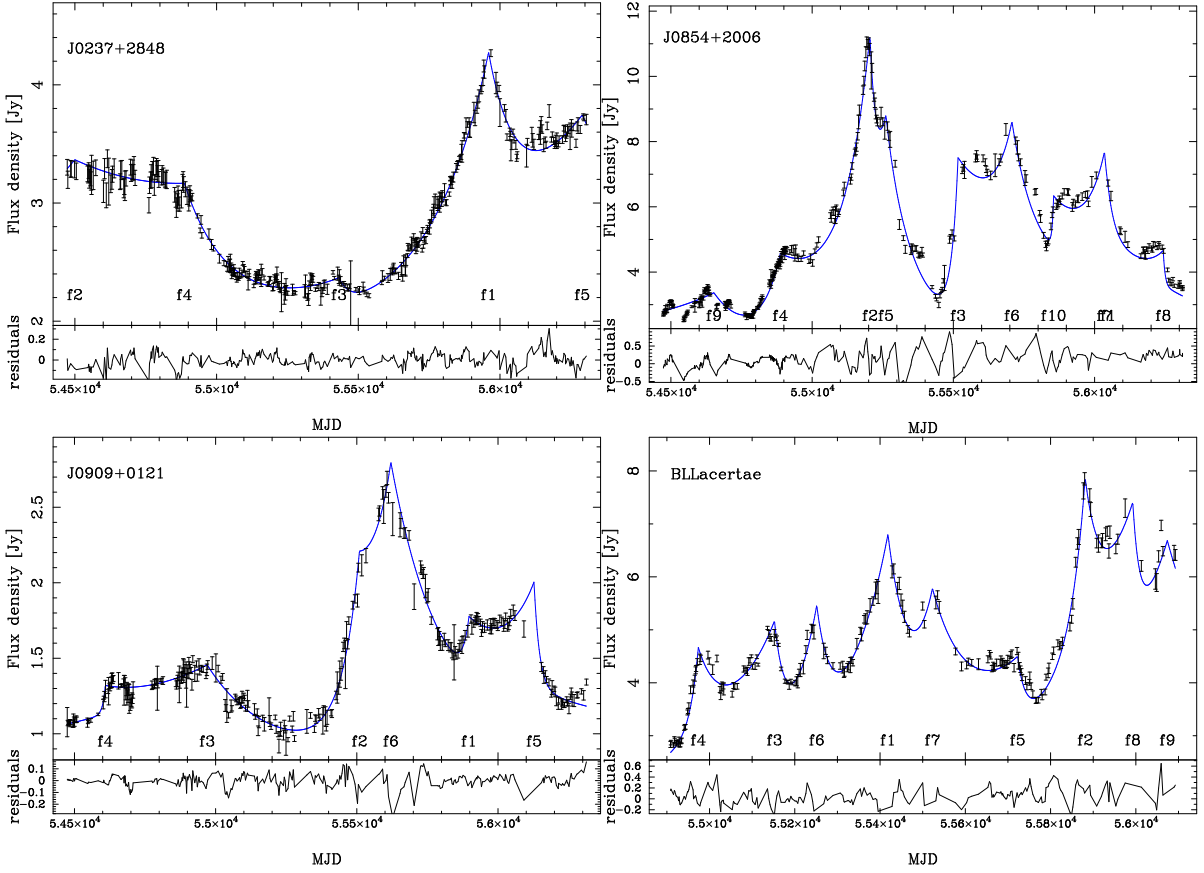

\centering
\includegraphics[scale=0.3,angle=-90]{J0237+2848_fit.ps}\includegraphics[scale=0.3,angle=-90]{J0854+2006_fit.ps}
\includegraphics[scale=0.3,angle=-90]{J0909+0121_fit.ps}\includegraphics[scale=0.3,angle=-90]{BLLacertae_fit.ps}
\caption{Examples of MCMC fits for four sources in the OVRO program. Top panel shows the data points and the median fit (blue curve) and the bottom panel shows the residuals of the fit. The source name is indicated in the top left corner of the plot. Each flare is labeled at the peak location of the flare.}
\label{fig:lc}       
\end{figure*}

Using 22\,GHz VLBI observations and total flux density variations, L\"ahteenm\"aki, Valtaoja \& Wiik showed that the typical values for intrinsic brightness temperatures are near the equipartition limit \cite{lahteenmaki99}. They adopted a value of $T_{\rm b,int} = 5 \times 10^{10}$\,K to estimate the Doppler boosting in a sample of 81 sources \cite{lahteenmaki99b}. The study was later expanded to include longer term observations and a larger sample \cite{hovatta09}. One of the main shortcomings of these studies is the lack of proper error estimates for the intrinsic brightness temperature and the Doppler boosting factor. We present here preliminary results from a study where we implement a Markov Chain Monte Carlo (MCMC) method to obtain the variability brightness temperatures, including proper error estimates. Very Long Baseline Array (VLBA) data from the MOJAVE (Monitoring of Jets in Active galactic nuclei with the VLBA Experiments) will be used to estimate the intrinsic brightness temperatures.

\section{Observed brightness temperatures}\label{sec-1}
We use the total flux density observations and light curves from the Owens Valley Radio Observatory (OVRO) blazar monitoring program \cite{richards11} to obtain  variability brightness temperatures at 15\,GHz. A sample of $\sim$1800 objects are monitored twice per week with the OVRO 40-m telescope to follow the changes in the total flux density. 
Blazar light curves can be modeled with flares of exponential rise and decay \cite{valtaoja99} so that the individual flares follow 
\begin{equation}
S(t) = \Delta S e^{(t-t_{max})/\tau},
\end{equation}
where $\Delta S$ is the amplitude of the flare, $t_{max}$ is the peak location of the flare and $\tau$ is the rise or decay time of the flare. These are the four unknown parameters for each flare. In addition, we fit for a baseline for each light curve, which adds one additional parameter.
By obtaining $\Delta S$ and $\tau$ we can estimate the variability brightness temperature as 
\begin{equation}\label{eq:Tvar}
T_{\rm var} = 1.548 \times 10^{-32} \frac{\Delta S d_L^2}{\nu^2\tau^2(1+z)},
\end{equation}
where $d_L$ is the luminosity distance to the object, and $z$ is the redshift \cite{hovatta09}. 

\subsection{Markov Chain Monte Carlo fitting of light curves}\label{sec-2}
Fitting is performed with the Climax codebase (Leitch et al., in
prep.), a generalized Monte Carlo framework for fast Markov Chain
sampling of the posterior likelihood.  The Climax code employs the
Metropolis-Hastings algorithm, with adaptive Hessian-based refinement
of the jumping distribution during the burn-in period to improve
convergence speed. The
code is heavily parallelized to take advantage of multi-processor
platforms. An object-oriented implementation makes it simple to add
extensions for new types of datasets and models, and to perform
simultaneous fitting of complementary datasets (the light-curve
fitting described here is a good example, as the code was originally
developed for joint fitting of X-ray and interferometric data).

The output from the fitting is a posterior distribution for each of the parameters from which we can determine the mean value and an error. The distribution is not necessarily Gaussian or even symmetric, which gives additional information on the complexity of the fit.
Examples of our light curve fits are shown in Fig.~\ref{fig:lc}. The curve in each plot is calculated by taking the median value of the parameters from three independent MCMC runs to verify the convergence of the fit to a global solution. We determine the brightness temperature of each flare by applying Eq.~\ref{eq:Tvar} to the posterior distributions of each parameters to obtain a {\it distribution} of $T_{\rm var}$. From this distribution we can estimate the most likely value and its error, in addition to being able to verify if the exponential curve is a good fit to the data. 

\begin{figure}[!ht]
\centering
\includegraphics[scale=0.5]{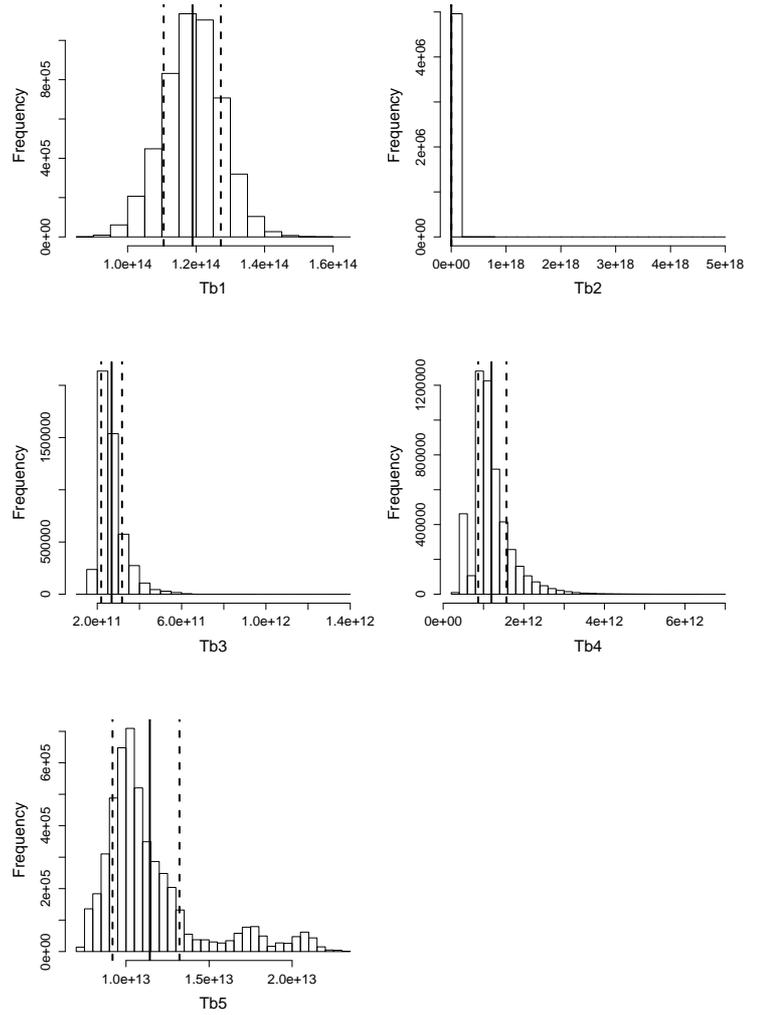}
\caption{Distribution of $T_{\rm var}$ for the flares in J0237+2848. The distributions are numbered as the flares in Fig.~\ref{fig:lc} top left panel. Black solid line shows the mean of the distribution and the black dashed lines the 68\% confidence limits.}
\label{fig:Tdist}       
\end{figure}

Examples of the distributions of $T_{\rm var}$ for the flares in source J0237+2848 (light curve in top left panel of Fig.~\ref{fig:lc}) are shown in Fig.~\ref{fig:Tdist}. It is clear that most of these are non-Gaussian and have a wide range of values making it difficult to determine the best-fit value. This is not surprising as most of the flares in J0237+2848 are not well constrained by the data. For example, the flare labeled f2 has no data to constrain the rise time, which is reflected in the wide range of $T_{\rm var}$ values (distribution Tb2). This can be used as additional information to select the robust estimates for $T_{\rm var}$. In the case of J0237+2848 we only use the $T_{\rm var}$ estimated from flare labeled f1 (Tb1) in our further analysis.

\subsection{Brightness temperatures from VLBA data}
We use 15\,GHz data from the MOJAVE project to study the VLBI brightness temperatures (Homan et al., in prep.). A sample of $\sim$300 objects have been regularly observed with the VLBA to monitor the changes in total intensity and polarization in the parsec-scale jets of AGN \cite{lister09a}. The basic procedure for obtaining the brightness temperature is to 
make an image of the source and identify the peak. In order to obtain an estimate of the brightness temperature in the core, all model components are deleted in the region surrounding the peak where the brightness is greater than 30\% of the peak. We then add a single Gaussian component to the peak location and fit the model to the data. A second Gaussian component is added to evaluate which model is better. We then calculate the brightness temperature, $T_{\rm VLBI}$, for the preferred model, and in the case of a model with two Gaussian components, use only the one closest to the peak brightness location. The outcome of this procedure is the core brightness temperature of the source at multiple epochs. 

\section{Intrinsic brightness temperatures}
Our preliminary sample includes 20 objects that have flared at least once since the beginning of the OVRO monitoring program in 2008. This sample is highly subjective, selected to test the fitting algorithm, and mainly includes sources with only a few flares. We emphasize that the results are preliminary.

\begin{figure}[!ht]
\centering
\includegraphics[scale=0.5]{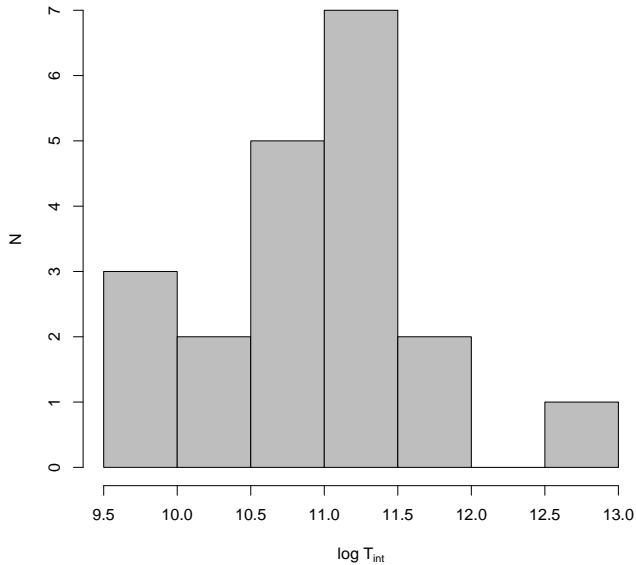}
\caption{Distribution of ${T_{\rm b,int}}$.}\label{fig:Tint}
\end{figure}
In order to calculate the intrinsic brightness temperatures, we need simultaneous total intensity and VLBI data. There are 20 flares (in 14 objects) for which there is a MOJAVE epoch within 2 months of the peak of the total intensity flare. This limit was selected because it ensures that the data were taken during the same flare. We calculate the intrinsic brightness temperatures for these flares using Eq.~\ref{eq:Tint}, and the distribution of values is shown in Fig.~\ref{fig:Tint}. Even with this preliminary and small sample, there is a clear peak at $T_{\rm b,int} \sim 10^{11}$\,K. This is close to the equipartition brightness temperature and in very good agreement with previous studies \cite{readhead94,lahteenmaki99,homan06}.

There are several caveats with the method that should be kept in mind when interpreting the results. Firstly, the variability brightness temperature definition assumes that the rise time of the flare corresponds to the light travel time across the source. If this is not the case, we will only obtain a lower limit of the brightness temperature. Secondly, the choice of an exponential flare shape is arbitrary and not based on physical modeling (unlike the model by M. F. Aller et al. these proceedings), and therefore does not always fit the data. This introduces problems in the fitting algorithm when a global solution does not exist or has a very broad distribution.

Our goal is to develop an automated way to fit a large number of flares to enable studies of all the flaring MOJAVE sources in the OVRO sample in order to determine the typical range of intrinsic brightness temperatures in parsec-scale blazar jets. Ultimately we will fit the light curves of all flaring OVRO sources to determine the Doppler boosting factors in a much larger sample than studied before.

\section*{Acknowledgments}
T.H. and J.R. acknowledge the support of the American Astronomical Society and the National Science Foundation in the form of an International Travel Grant. T.H. was supported by the Jenny and Antti Wihuri foundation. The OVRO 40-m monitoring program is
supported in part by NASA grants NNX08AW31G 
and NNX11A043G, and NSF grants AST-0808050 
and AST-1109911. The MOJAVE project is supported under NASA-{\it Fermi} grants NNX08AV67G and 11-Fermi-11-0019.

%
\bibliography{thbib}

\begin{thebibliography}{9}

\bibitem{kellermann69}
K.I. {Kellermann}, I.I.K. {Pauliny-Toth}, \apj{} \textbf{155}, L71 (1969)

\bibitem{readhead94}
A.C.S. {Readhead}, \apj{} \textbf{426}, 51 (1994)

\bibitem{lahteenmaki99}
A.~L\"ahteenm\"aki, E.~Valtaoja, K.~Wiik, \apj{} \textbf{511}, 112 (1999)

\bibitem{lahteenmaki99b}
A.~L\"ahteenm\"aki, E.~Valtaoja, \apj{} \textbf{521}, 493 (1999)

\bibitem{hovatta09}
T.~{Hovatta}, E.~{Valtaoja}, M.~{Tornikoski}, A.~{L{\"a}hteenm{\"a}ki}, \aap{}
  \textbf{494}, 527 (2009)

\bibitem{richards11}
J.L. {Richards}, W.~{Max-Moerbeck}, V.~{Pavlidou}, et~al., \apj{}s \textbf{194}, 29 (2011)

\bibitem{valtaoja99}
E.~Valtaoja, A.~L\"ahteenm\"aki, H.~Ter\"asranta, M.~Lainela, \apj{}s
  \textbf{120}, 95 (1999)

\bibitem{lister09a}
M.L. {Lister}, H.D. {Aller}, M.F. {Aller}, et~al., \aj{} \textbf{137}, 3718 (2009)

\bibitem{homan06}
D.C. {Homan}, Y.Y. {Kovalev}, M.L. {Lister}, et~al., \apjl{} \textbf{642},
  L115 (2006)

\end{thebibliography}
%
%
%
%

\end{document}